# URL ORDERING POLICIES FOR DISTRIBUTED CRAWLERS: A REVIEW

**Deepika**
Assistant Professor,
*Computer Engineering Department,*
*YMCAUST, Faridabad 121006*
Email: deepikapunj@gmail.com

**Dr Ashutosh Dixit**
Associate Professor,
*Computer Engineering Department,*
*YMCAUST, Faridabad 121006*
Email:dixit_ashutosh@rediffmail.com

*Abstract*—With the increase in size of web, the information is also spreading at large scale. Search Engines are the medium to access this information. Crawler is the module of search engine which is responsible for download the web pages. In order to download the fresh information and get the database rich, crawler should crawl the web in some order. This is called as ordering of URLs. URL ordering should be done in efficient and effective manner in order to crawl the web in proficient manner. In this paper, a survey is done on some existing methods of URL ordering and at the end of this paper comparison is also carried out among them.

## I. INTRODUCTION

As the size of web increases, it is necessary for search engines to enrich their databases with fresh and latest information. Information on the web is in the form of web pages. So, there should be some way to get fresh pages in search engines database. The crawler is the module which is responsible for downloading web pages from web. It starts with a URL from seed URLs list and downloads the web pages. It also extracts URLs embedded there in and adds these URLs in the URLs queue and so on. This process of crawling should be optimized in a way to cover maximum size of web. There are many design issues related to design of a crawler [1]. Each issue has its own role in order to work crawler proficiently. If the work of crawling is either parallelized or distributed by creating crawler instances, say agents. The general architecture of Distributed crawling is shown in figure 1. But by creating agents, many other problems may arise. One of the major problems is the problem of duplicate downloading of URLs. One URL can be downloaded by multiple agents and thus wasting bandwidth and network resources.

As shown in figure 1, crawler has multiple agents that are distributed over World Wide Web. These agents crawl the web and send the pages to the crawler. Crawler stores the pages and internal links are extracted from them and added to Queue. Form Queue URLs are submitted again to crawler and then crawler sends them to agents and the whole process repeats.

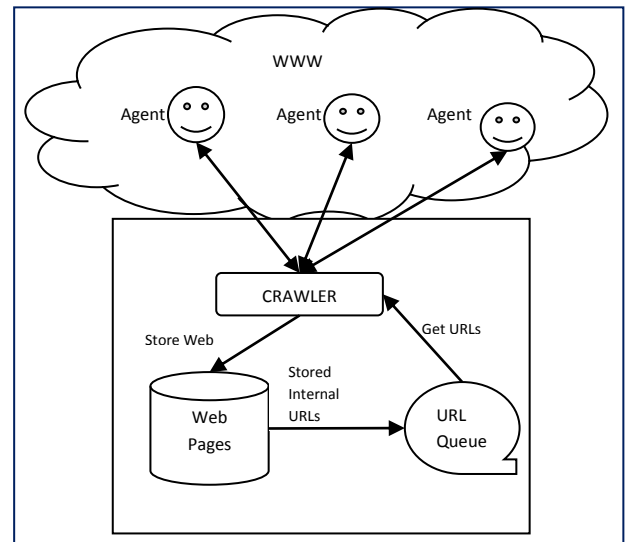

Figure 1: General Architecture of Distributed Crawler

## II. URL ORDERING

Ordering of URLs is an important concern for efficient downloading of web pages. By Ordering, crawler agents will get list of unique URLs and thus parallelize the downloading in appropriate way. To do this job in proper manner, following issues should be taken care of:

- ➢ Check of duplicity
  When crawling is carried out in parallel then there are chances of downloading or accessing the same URL by different agents.

- ➢ Network Resources Utilization
  There should be proper utilization of network resources. It should not be the case that some agents are utilizing resources more and some are waiting for long.
- ➢ Load Balancing
  Load should be properly distributed to the agents. All agents get the load equally. It should not be the case that some are sitting idle while others are overloaded.

### III. URL ORDERING POLICIES

Many researchers have done work in this area. Some of their work is discussed below with their advantages and disadvantages.

*A. Task Based Scheduling*

Dajie et al [2] proposed a URL scheduling algorithm based on Round Robin Scheduling. They used weight factor to schedule the URLs. For calculation of weight, they used time as important factor. Time value more means crawler has tasks that are yet not completed. It means it should not get more tasks. They take weight and time as reciprocal of each other. So, weight is low for that node which has more time to finish its task. It uses master slave architecture for scheduling purpose. The flow of this master slave is shown in figure 2.

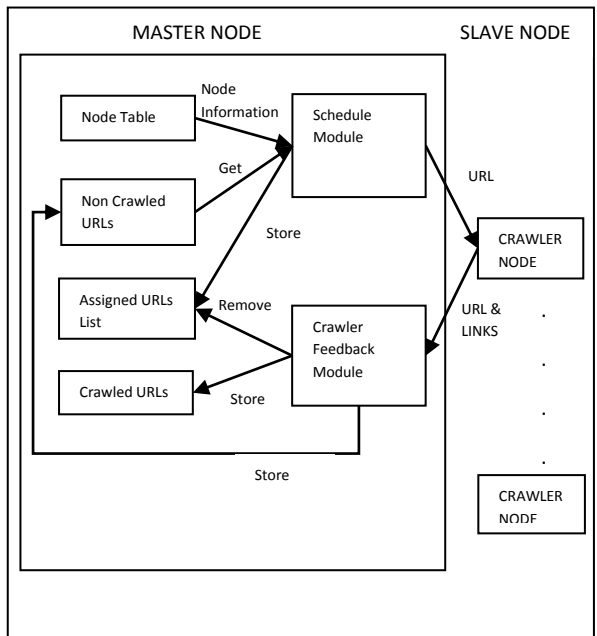

Figure 2: Master Slave Architecture of URL Scheduling

In master slave architecture, at master node various data structures are used to store the information of crawler node as well as status of URLs i.e. whether they are scheduled to be crawled or completely crawled. They used the concept of Round Robin algorithm for assigning URLs to crawler. A weight is assigned to each crawler node and hoping that weight will show the status of that node. If weight is low then it means crawler is heavily loaded and vice versa. The weight is assigned with the help of given relation:

$$W = \frac{k}{\sum_{i=1}^{k} t_i * (m + 1)}$$

Where, k=no. of tasks finished recently
$t_i$ =finished time of i tasks
m=no. of tasks yet not finished

With the help of this weight value scheduling is done. URLs get scheduled on the basis of this weight value of crawler node. A threshold value is taken to ensure that low weight crawler node will not leave unattended.

Advantages
- ➢ It is simple and efficient:-It works at master node and takes less time. Due to fast nature of Round Robin Scheduling, it prevents the crawler from sitting idle. It takes less time for scheduling also.
- ➢ Supports dynamic entry of crawler node:- It has been observed in existing algorithm, on new entry whole algorithm needs to restart. But in this algorithm, new entry doesn't effects the working and also taken into consideration at that point of time.
- ➢ No starvation:-The node with lower weight doesn't leave idle in this algorithm. As it has been observed that existing algorithm suffers from this starvation problem. In this algorithm, all crawler nodes get the chance to do crawling instead of its low weight.
- ➢ Error Recovery mechanism:-If some node doesn't respond or crawler node may crash, then there is recovery mechanism is there in this algorithm.

Disadvantages
- ➢ Single point of failure:-If master node crashes then all information of nodes gets lost. In this algorithm, there is no mechanism for master node recovery.
- ➢ Scalability:-If number of URLs increases then whether this algorithm works efficiently or not is not taken into consideration.

*B. Hashing Algorithm*

Yuan Wan et al [3] designed and implemented a URL assignment method based on hashing. It works on parallel systems [4] where systems are physically independent but they are cooperating with each other through some mechanism. Each system downloads the webpages on their local machine and when internal links are extracted from these webpages, there is need of scheduling of these URLs. Either they are scheduled to be downloaded on host machine or to some other machine. The crawlers are not communicated with each other. Host Machine is the central coordinator through which communication and scheduling takes place. The architecture of the system is as shown in figure 3.

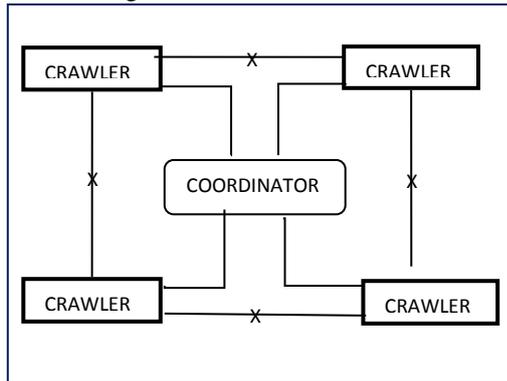

Figure 3: Model of Distributed Parallel Crawler

Coordinator assigns the URLs to different crawler based on its host name. Host name decides that whether the URL goes to other machine or downloads on its home machine. For this purpose a hashing scheduling algorithm is designed. It will take URL as input and then hash function will apply on this URL. In hashing function, host name is extracted from URL then convert it into integer format and will match with id of crawler. If match takes place, then it will download on same machine otherwise will go to other. The coordinator has id of all registered crawler and on the basis of this information, it will schedule the URLs. The step wise execution of this algorithm is as follows:

*Step 1: Each crawler get registered to coordinator and get an ID which is in integer form.*
 *2: Apply hash function to each URL.*
  *2.1 $Key_i = (\sum_{i=1}^{l} Transfer(host(URLi)) \bmod n$*
// Here, transform convert the host into integer
 *3: At coordinator, if (key==ID)*
  *3.1 If (URL exists in the URL_List), then*
   *3.1.1 Save the page*
 *4: Else*
  *4.1 save the URL*
  *4.2 Send to coordinator*
 *5: End*

Advantages
- It prevents duplicate URLs to download again. It partitioned the URLs list in such a way that ensures that no URL repeated at any machine.
- By preventing repeated access of same URL at different machine also saving bandwidth and network resources very well.

Disadvantages
- It is not scalable i.e. if number of URLs increases then it is not sure that they performed in same manner as it does now.
- It has single point of failure. It works according to coordinator directions and if coordinator crashes whole algorithm goes down and system stops working.

*C. Popularity based*

Chandramouli et al [5] proposed the URL ordering based on popularity. In this technique, web logs available on website were used for calculating total access counts for each URL. URL ordering were classify into two approaches. One is non learning algorithms that uses predetermined ordering function and other is learning algorithms that will orders the URLs based on training set of URLs with quality information.

In non-learning algorithm, high the access count the more important is the page. But, it is also suggested that it may case that website owner itself access its website several times and this cause increase in access count and considered as important page. To avoid such situation, four types of accesses to a website were considered. These are:

1. Total External Count:- access to URL on website from outside the local network
2. Unique External Count:- unique access from outside the local network
3. Total Internal Count:- access to URL on website from local network
4. Unique Internal Count:- unique access from local network

Thus, by calculating all above access count by using the given below relation, total access count is obtained.

Total= TEC+UEC+TIC+UIC

To predict the importance of every count value their accuracy is calculated and then assigns weights to each count value with the help of these accuracy values. Thus weighted score of each URL is calculated as follows:

WeightScore=a*$TEC_{acc}$/Total+b*$UEC_{acc}$/Total+c* $TIC_{acc}$/Total+d*$UIC_{acc}$/Total

Where, $TEC_{acc}$ =TEC accuracy algorithm
$UEC_{acc}$ =UEC accuracy algorithm
$TIC_{acc}$  = TIC accuracy algorithm
$UIC_{acc}$  = UIC accuracy algorithm

a, b, c, d are raw external, unique external, internal and unique internal counts for the URL.

In learning algorithm, best combination of above four count values was used. Two learning algorithms were implemented, Total Access Count Learning (TAC-L) and Split Access Count Learning (SAC-L). Both algorithms have training and testing phases. In these algorithms, access counts as input and supplied to any learning algorithm like decision tree or k-nearest neighbour and model is prepared. To measure the quality of a page, Page Rank algorithm was used. Higher the rank, more important is the page. The working of learning algorithm is shown in figure 4.

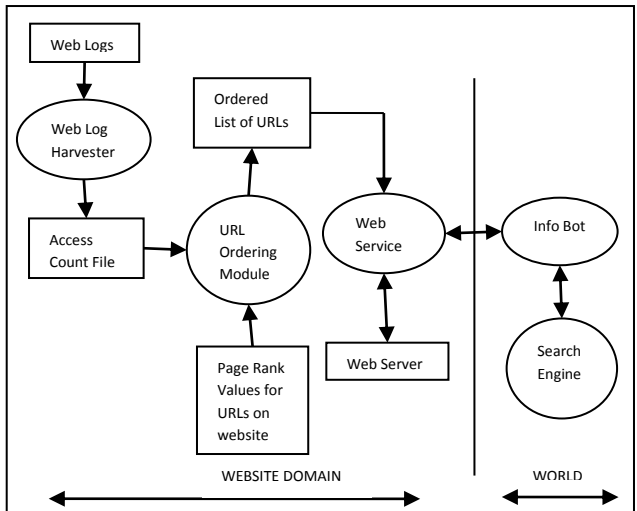

Figure 4: Popularity Based Architecture of Search System

Here, Web log server files are used to calculate the access counts of each URL and supplied as input to learning algorithm. With the help of Page Rank algorithm, URL can be ordered and provide to Info Bot for crawling.

Advantages
- It works on popularity of URLs, it means it considers users' interest while ordering the URLs.
- It put less burdens on search engine as each website maintains its own log file.
- It is better than BFS. It doesn't miss important pages which are in deeper depth of websites.

Disadvantages
- It works on small data set.
- As it considers popularity factor for URL ordering so it may case that this factor may be manipulated.
- It doesn't work on new pages which are added later in the website.

*D. Dynamic URL Assignment*
A.Guerriero et al [6] proposed a dynamic URL assignment method based on fuzzy clustering. It worked on the principle that making clustering of URLs and then scheduled them. Clustering should be done in such a way that same URL shouldn't be crawled by multiple crawlers. Assignment should be done in such a way that unique crawling is done and it should be in optimizing manner.

A distributed architecture of parallel web crawler was proposed. Its components cooperate in efficient manner in order to get desired results. The architecture is shown in figure 5.

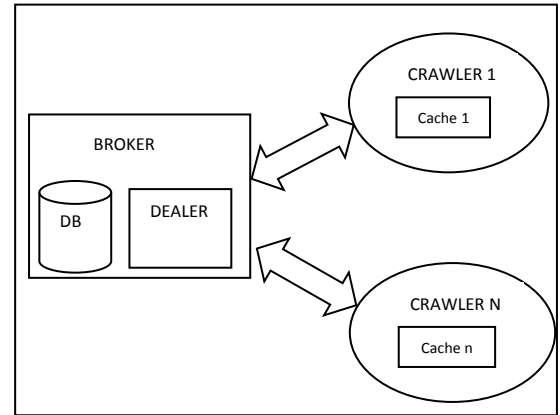

Figure 5: Dynamic Assignment of URL model

It consists of following components:-
1. Broker
   It is responsible for scheduling of URLs and also to create communication link between crawler and database. It picks the URL form database and with the help of dealer scheduled it to crawlers.
2. Dealer
   It is responsible for crawler efficient working. It manages load optimization with the help of fuzzy clustering method.
3. Crawlers
   It is the module which is responsible of crawling the web and downloads the page in database.

Following are the steps of working:
Step 1: *Get the URL form database.*
    2: *Divide the URL structure into different components as mentioned by URI standards [7].*
    3: *Apply Hash function to each component and convert it into integer form.*
    4: *Represent integer URL components into 3D coordinates.*
    5: *Apply Fuzzy Clustering [8] to these URLs.*
    6: *Assign these clusters to different crawlers for crawling.*
    7: *Extracted links checks for duplicity with already stored in database.*
        7.1 *If already found in database, then*
            7.1.1 *Discard*
            Else
            7.1.2 *Stored in database.*
    8: *Go to Step 1.*
    9: *End.*

Advantages:
- It checks for duplicity. It means prevents downloading of same URLs again & again and thus prevents bandwidth.
- It maintains load on the system efficiently by assigning clusters to crawlers.

Disadvantages
- Communication Link failure. It highly depends upon communication for its working and if link lost whole algorithm fails.
- Single point of failure. One system is treated as broker and broker initiates the algorithm. So, if this system crashes, broker lost and algorithm fails.

## IV. Conclusion

Crawler plays an important role in gathering data and stored in database. It is the responsibility and objectives of crawler to maintain database rich and fresh. In order to achieve these two objectives i.e. rich and fresh, crawler has to order URLs. This URL Ordering helps the crawler to maintain the richness and freshness of database. A comparison table of different methods of URL ordering is shown below:

Table 1: Comparison of URL ordering

| Algorithm | Task Scheduling | URL Hash | Popularity Based | Dynamic URL |
|---|---|---|---|---|
| Technique Used | Round Robin | Hashing | Links Count | Fuzzy clustering |
| Description | Works on round robin policy i.e. every crawler get its turn and weight assigned get surety of load balancing | Hashing of URL is done. In hashing method host is find out and then convert the characters into its integer equivalent and matched with crawler's ID. | External and internal links count is calculated. | Clustering is done to schedule URLs. |
| Input | Time | URL | Link Count | URL |
| Resource Utilization | Efficient | Efficient | No consideration | Efficient |
| Duplicity Check | Yes | Yes | No | Yes |
| Recovery | Yes | Not Considered | Not applicable | No |
| Advantages | Simple & efficient, no starvation, load balancing | Fast, Load balancing | Simple and better than BFS, less burden on search engines | Efficient, Load balancing, remove duplicates, low cost |
| Disadvantages | Scalability, Single point of failure | Lack of scalability, less realistic, single point of failure | Small data set, not applicable to new pages | Dependency on communication link, single point of failure |

In task scheduling time is taken as scheduling criteria that is less time taken URLs will be schedule first while popularity based took number of links on that URL as scheduling criteria. URL Hash and Dynamic URL both took URL itself as scheduling criteria. They work on URL and then decide their order for crawl.

## REFERENCES

[1] Deepika, Dixit Ashutosh, "Web Crawler Design Issues: A Review", published in International Journals of Multidisciplinary research Academy (IJMRA) in August 2012.
[2] Dajie Ge , Zhijun Ding , 'A Task Scheduling Strategy based on Weighted Round-Robin For Distributed Crawler', IEEE/ACM 7th International Conference on Utility and Cloud Computing, 2014.
[3] Y. Wan, H. Tong. – "URL Assignment Algorithm of Crawler in Distributed System Based on Hash". - IEEE International Conference on Networking, Sensing and Control, ICNSC 2008, Hainan, China, 6-8 April 2008. pages 1632-1635, IEEE, 2008.
[4] M. Marin, R. Paredes and C. Bonacic, "High-performance priority queues for parallel crawlers," Proceedings of the 10th ACM workshop on Web

information and data management, October 2008, pp. 47-54.

[5] Arvind chandramouli , Susan gauch and Josua eno "A popularity-based URL ordering Algorithm for Crawlers", M*ay 13-15 2010"* 3rd Conference on Human System.

[6] A Guerriero, F. Ragni, C. Martines, "A dynamic URL assignment method for parallel web crawler", IEEE International Conference, 2010.

[7] http://tools.ietf.org/html/rfc3986 - RFC 3986 / STD 66 (2005) – the current generic URI syntax specification.

[8] Applying Preference-based Fuzzy Clustering Technology to Improve Grid Resources Selection Performance - Dong Guo, Liang Hu, Shilan Jin and Bingxin Guo - Fourth International Conference on Fuzzy Systems and Knowledge Discovery (FSKD 2007).

[9] LI Xiao-Ming, and FENG Wang-Sen. "Two Effective Functions on Hashing URL" Journal of Software, 2004.

[10] B. Debnath, S. Sengupta, J. Li, D.J. Lilja, and D.H.C. Du, "BloomFlash:Bloom filter on flash-based storage," 2011 31st International Conference on Distributed Computing Systems, June 2011, pp. 635-644.

[11] Christopher Olston, Marc Najork "Web Crawling", Foundations and Trends in Information RetrievalVol. 4, NO. 3 (2010)

[12] Wani Rohit Bhaginath, Sandip Shingade, Mahesh Shirole, "Virtualized Dynamic URL Assignment Web Crawling Model", IEEE International Conference on Advances in Engineering & Technology Research (ICAETR - 2014) 2014.